\documentclass[twocolumn,showpacs,prl]{revtex4-1}
\usepackage{amsmath}
\usepackage{graphicx}
\usepackage{todonotes}

\usepackage{color}
\usepackage{algorithm2e}

\begin{document}
\vspace*{0.35in}

\title{Selection first path to the origin of life}

\author{Nicholas Guttenberg, Nathaniel Virgo, Chris Butch, Norman Packard}
\affiliation{Earth-Life Science Institute, Tokyo Institute of Technology, Meguro, Tokyo 152-8551, Japan}

\begin{abstract}
We propose an alternative to the prevailing two origin of life narratives, one based on a replicator first hypothesis, and one based on a metabolism first hypothesis.  
Both hypotheses have known difficulties:  All known evolvable molecular replicators such as RNA require complex chemical (enzymatic) machinery for the replication process.  Likewise, contemporary cellular metabolisms require several enzymatically catalyzed steps, and it is difficult to identify a non-enzymatic path to their realization. We propose that there must have been precursors to both replication and metabolism that enable a form of selection to take place through action of simple chemical and physical processes.  We model a concrete example of such a process, repeated sequestration of binary molecular combinations after exposure to an environment with a broad distribution of chemical components, as might be realized experimentally in in a repeated wet-dry cycle.  We show that the repeated sequestration dynamics results in a selective amplification of a very small subset of molecular species present in the environment, thus providing a candidate primordial selection process.

\end{abstract}

\maketitle

\section{Introduction}

The chemistry of life has a number of unique properties in comparison to other chemical systems. In principle, the chemistry of a cell can produce any sequence of amino acids, and therefore any compound that can be synthesised using proteins. There are a number of ``messy'' prebiotic chemistries that can also produce a combinatorial explosion of diverse products, such as Miller-Urey synthesis, HCN polymerisation, formose chemistry, Fisher-Tropsch-type reactions and others. However, life exhibits a great deal of self-restraint in the molecules it produces. It samples only a tiny portion of its combinatorial space, generally only producing molecules that are useful to it, and molecules closely related to them. 

We consider a class of physical processes acting on large libraries of molecules to cause a selective amplification of small subsets of the library.  The processes tend to sequester molecules pairwise into a reservoir, based on the strength of pairwise interaction between the molecules.  One version of such a process is precipitation of molecular combinations from evaporation, forming a reservoir of precipitants.  We present a simple model for the process that has sequestration rates for the pairs given by a Maxwell-Boltzmann distribution of the pairwise binding strengths.  Repeated sequestration governed by the dynamics of this model can result in selective amplification of a subset of molecules, if the temperature is cool enough. We quantify this selective amplification and study its dependence on system parameters.  Finally, we discuss the role that selective amplification might play in the origin of life, quite possibly preceding molecular replication as a source of informational persistence and variation in primordial evolutionary processes.

The production of large, complex libraries of chemical compounds could play an important step in the origin of life. However, a large complex library is not enough, in itself, to produce life; ``messy'' prebiotic chemistries may produce biological ``building blocks'' among other molecules, but they lack the emergence of intricate functionality characteristic of life.  Extracting interesting functional subsets of functional molecules from such libraries then becomes a hurdle for the origin of life.  In experiments, this hurdle is routinely overcome by human intervention in choosing the target of selection, using methods such as binding assays to search for compounds with useful or interesting binding properties. Emergence of interesting functional subsets at the origins of life would require such systems to perform selection on themselves --- perhaps even before the emergence of things such as information-carrying polymers and replicators could occur. Otherwise, even if such functional molecules existed somewhere within the ensemble, the presence of a large quantity of inert compounds and/or an abundance of parasitic side reactions would inhibit the emergence of complex functionality.

Attempts to resolve this issue have primarily centered around looking for factors which would drive the chemistry itself to produce fewer things, but in higher yields. For example, the addition of borate salts prevents tar production in the formose reaction\cite{benner2010planetary}. However, as a result the chemistry loses some ability to search the chemical space on its own --- when fewer things are made, there is less chance of one of those things happening to be a particular useful catalyst or functional molecule.  Additionally, optimizing the yield generally requires some control over the reaction conditions --- not just including some materials in the reaction vessel, but also excluding contaminants --- which sharply limits the types of compatible environments.

Instead, is there a way for chemical sparseness to emerge on its own, not as a result of controlling the underlying chemical reaction network itself but as a result of physical processes acting upon the complex library produced by an uncontrolled chemistry? This can be done, for example, by performing an assay that selects only those materials that bind to some particular surface. There plenty of natural materials that could separate out compounds by affinity to the material. However, this will select for something that is relative to the context of the porous material, rather than selecting for something that relates to things such as chemical function and interactions. Instead, it seems necessary for the chemistry to find a way to perform a sort of binding assay upon itself, such that the properties that are selected for are themselves an emergent consequence of the compounds that are there. That it to say, we would need a way for there to be a positive feedback of selection upon itself, so that a chemical system which initially lacked a mechanism of selection and sparseness could spontaneously have one emerge and become locked into place.

When thinking selection in this sense, it is natural to divide the system into a portion which is retained and a portion which is discarded. The retained version acts as a sort of protected reservoir for materials with high affinity, while the fluid which is now depleted in those high-affinity materials flows out of the system and is replaced with fresh material. That is to say, the transport properties of materials in a chromatography column, for example, are dependent on their affinity, and that difference in transport properties is responsible for creating the increase in concentration of the high-affinity compounds within the column. In normal chromatography, the ability of a compound to enter the protected reservoir is a function of its affinity for the column material. In our case, we want the ability of a compound to enter the reservoir to instead be primarily dependent on the distribution of other compounds present.

For this purpose, we turn to wet-dry cycling. Recently, wet-dry cycling has received a lot of attention for its ability to promote polymerization\cite{mamajanov2014ester, rodriguez2015formation}. Our focus however is not on the tendency of drying to promote polymerization per se, but that during a drying cycle, compounds are both joining together into polymers and complexes as well as precipitating out of solution in order as a consequence of their ability to form polymers and complexes with other compounds currently in the mixture. That is to say, the system fractionates based on a property which exists because of the pairwise interactions between compounds rather than just an intrinsic property of each compound on its own.  A simple example of this phenomenon is in the pairwise precipitation of melamine or triaminopyrimidine and barbituric acid which are individually soluble but co-precipitate at moderate concentrations~\cite{seto1993molecular,cafferty2013efficient}. In a more complex system, nonlinear interaction of components during fractionation can provide the necessary ingredient to produce the type of positive feedback which might lead to the spontaneous emergence of selectivity and sparseness. 

To test these ideas, we will construct a general model of a wet-dry process without chemical transformation --- just reversible precipitation. The precipitation process selects for affinities between pairs of compounds, but because the overall concentration profile changes over time during the drying phase, those affinities also change in a complex, population-dependent way. We will show that when the pairwise interactions are sufficiently strong, and when the fraction of supernatant discarded is sufficiently small, this model has a phase transition from a complex mixture to a case where a sparse subset of the supplied compounds are amplified to much higher concentrations than the rest. The system sustains this sparse distribution despite having a constant influx of the chemically complex solution. 

Furthermore, the system is highly sensitive to small variations in the environmental availability of compounds. A small perturbation to the environment can sometimes lead to a disproportionately large change in the final concentration profile. This means that the system tends to amplify the diversity of any set of environments it operates in --- small local variations in concentrations or fluxes become large variations in the distribution of chemical repertoires. As such, this kind of emergent selection could help to explore chemical space in a pre-evolutionary world.

Wet-dry cycling may not be plausible in some prebiotic scenarios. However, this mechanism of transport to and from a protected reservoir to produce chemical sparseness is quite general. Other processes may be able to provide this as well, such as freeze-thaw cycles or cyclic flow through porous materials (which compounds which form larger complexes become trapped, while small molecules can more freely move back and forth). It may also be that in a spatially structured tar exposed to flow, differential viscosities between chemical complexes would be sufficient to enact this sort of effect.  The essential aspect that all these processes have in common is a repeated sequestration process, governed by a nonlinear pairwise interaction of the constituents, and as such, all of these processes may effectively be modeled by the dynamics presented below.

\section{Model}

\begin{figure}
\includegraphics[width=\columnwidth]{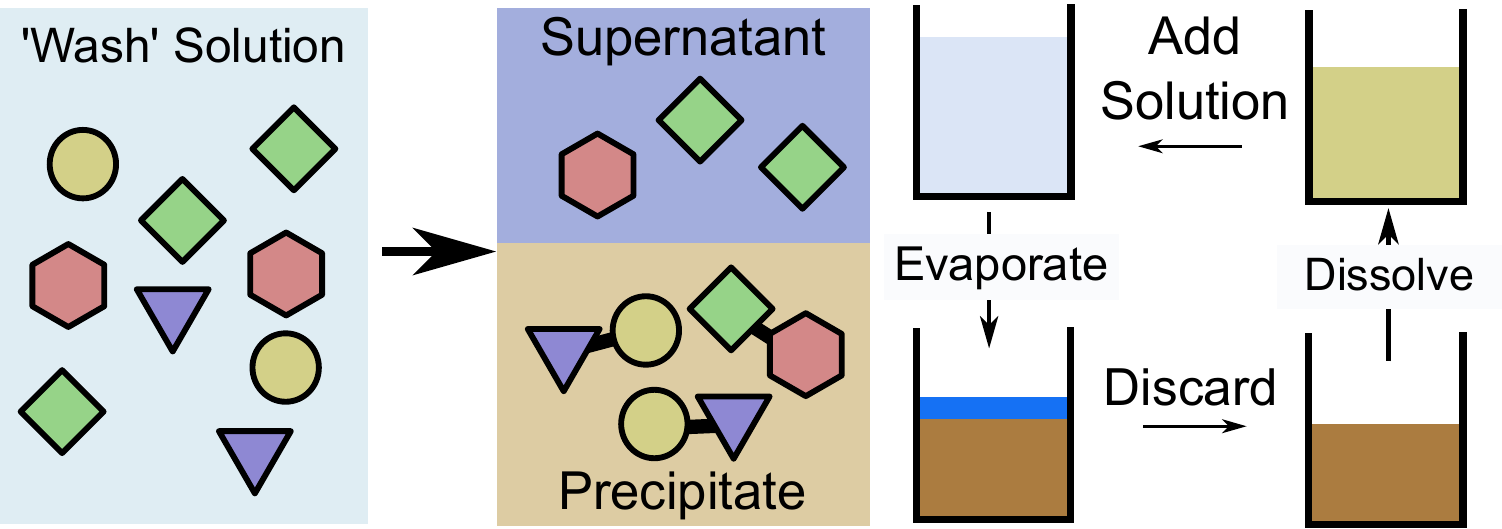}
\caption{Cycle of drying and wetting in the reservoir heredity model.}\label{ModelDiagram}
\end{figure}

We consider a pool of liquid in which $N$ types of compounds are dissolved, at concentrations described by the concentration vector $\vec{c}$. In addition, there is an environmental source of material with a vector of concentrations $\vec{e}$. As the liquid in the pool evaporates, the dissolved compounds precipitate out and form a solid. That solid is the 'reservoir', preserving the contained compounds until the next cycle. Compounds that do not make it into the reservoir are discarded and are replaced with fresh fluid from the environment, adding in new material and dissolving the existing precipitate (Fig.~\ref{ModelDiagram}). As a result, the profile of concentrations of compounds in the system gradually changes, with compounds that easily co-precipitate being retained at a higher rate than compounds which do not bind to anything currently in the system.

The dissolving stage is assumed to proceed to completion, so the only phase we need to explicitly model is the evaporation stage of the cycle. We specifically consider the case of rapid drying, so that precipitation is effectively irreversible. What matters then is the rate constants for the different pairwise binding processes. We model these rate constants by assigning each pair of compounds a transition state energy, which we specify by way of a symmetric matrix $J$, such that $J_{ij} = J_{ji}$ is the energy of the transition state between a compound of type $i$ and a compound of type $j$. The rate constant for the binding of $i$ and $j$ is then $e^{-\beta J_{ij}}$. We generally consider the behavior of random $J$ matrices, where each component of $J$ is sampled from a uniform random distribution $[-1,1]$; however, known rate information could equally well be used to map this model onto a specific chemical system.

The result is that we obtain a set of differential equations for the solution and reservoir concentrations, such that compounds $i$ and $j$ precipitate out together into the reservoir pair $r_{ij}$ according to:

\begin{equation}\label{Model}
\frac{d r_{ij}}{dt} = c_i c_j e^{-\beta J_{ij}}
\end{equation}

where $\beta$ is the inverse of the temperature.

This process proceeds until all but a fixed fraction of the compounds have entered the reservoir. This is the supernatant fraction $f$. We then discard the residual supernatant and re-dissolve the precipitate. Finally, because this process has reduced the total concentration of the compounds in the system, we add in new mass according to the environmental concentration vector $\vec{e}$ to bring the concentration back up to the initial value. The result is that at each cycle of the system, only the relative composition of the solution can change. The sum over the concentration vector is conserved: $\sum_i c_i = 1$.  We perform multiple cycles of this process (typically 4000 cycles, in the simulations below), and examine how the concentration vector of the system $\vec{c}$ changes over time.

Direct simulation of these differential equations is problematic, because there are large differences in reaction rates and the fast reactions will tend to finish before the target supernatant fraction has been reached --- effectively, this is a very stiff system. We address this issue by using a dynamic timestep size. At each point, some concentrations might become negative in the next timestep due to overshooting for a given timestep size. We determine the largest timestep size $\Delta_0$ such that no concentration will overshoot and become negative. If $\Delta_0$ is taken to be the timestep size, then each timestep at least one compound would have its concentration in the solution go from non-zero to zero. However, this can lead to noticeable discreteness effects, such as spurious oscillations. As such, we can in general use a fixed fraction of that timestep size to advance the system (usually $0.5\Delta_0$) to increase the stability of the simulation when needed.

\section{Results}

This section summarizes the results of numerical simulations of the model, starting from the general range of behaviors exhibited by the model and then proceeding to specific simulations to test whether the dynamics of this system correspond to the behavior of evolving population. 

First, we simulate the model for different values of the supernatant fraction and the temperature, to see if and when the sort of amplification mechanism we anticipated occurs. For a given distribution of concentrations, we need to characterize the degree of amplification which has occurred. Specifically, we are interested in regions of the phase diagram in which the distribution of compounds becomes sparser.  

\begin{figure}
\includegraphics[width=\columnwidth]{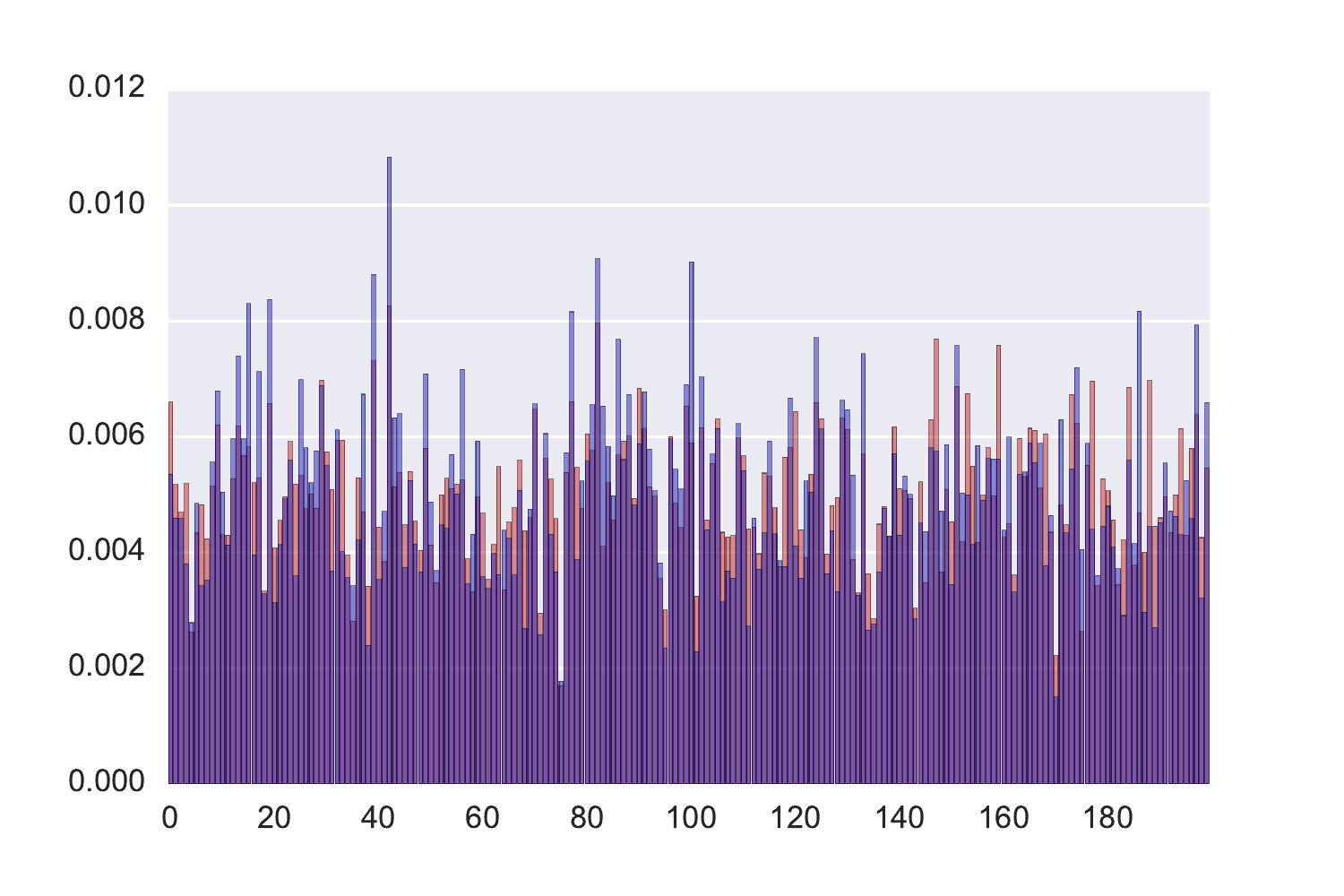}
\includegraphics[width=\columnwidth]{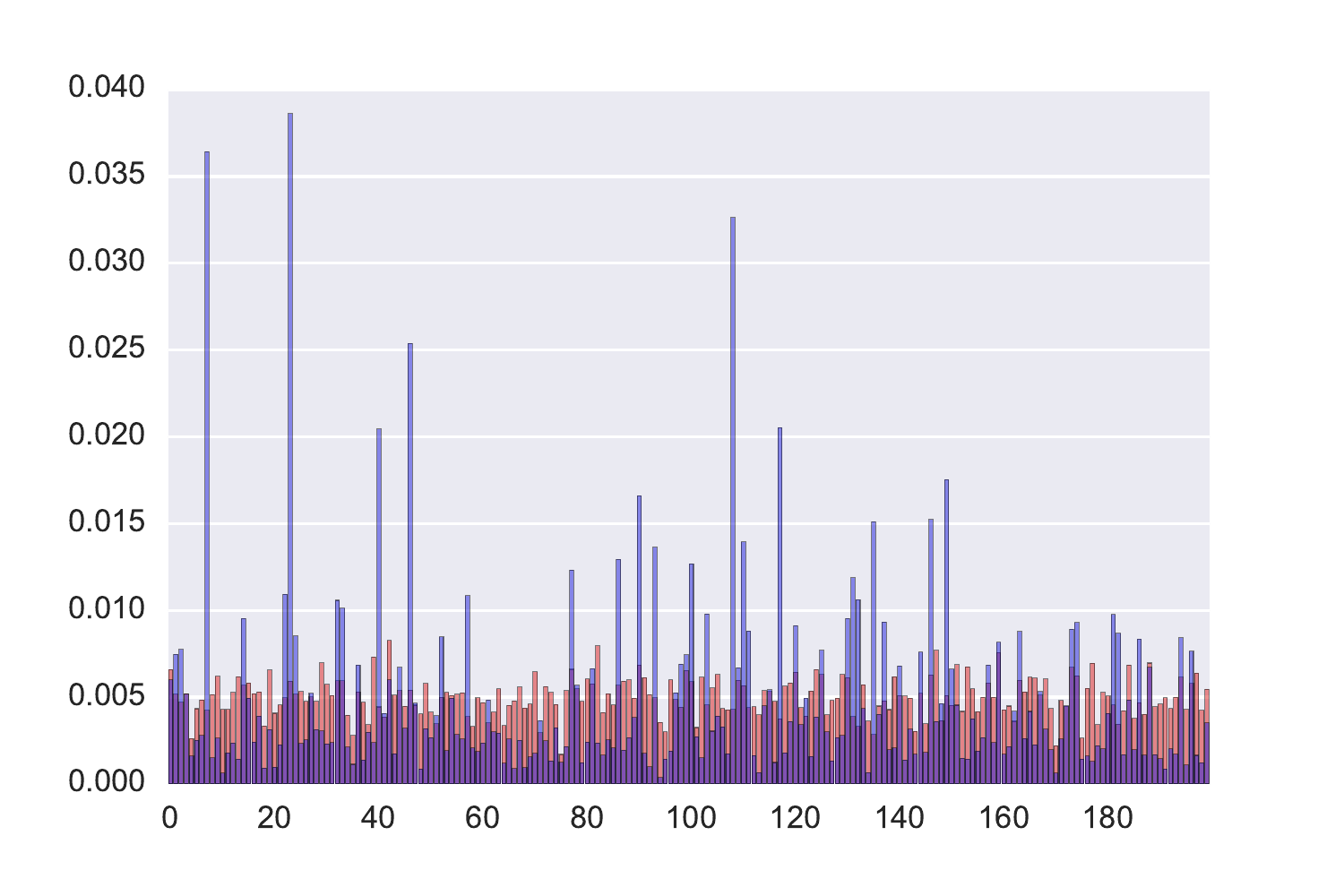}
\caption{Initial and final states, for high temperature ($\beta = 1$) and low temperature ($\beta = 100$).  Note that in the low temperature regime, a small subset of the peaks are selectively amplified by the nonlinear dynamics of repeated sequestration.  When selective amplification is observed, the non-amplified chemical species are suppressed, since the concentrations must add to one.
}\label{ConcentrationState}
\end{figure}

The basic phase transition is illustrated by comparing final states for two values of $\beta$, one below the phase transition, and one above the phase transition.  These two cases are illustrated in Fig.~\ref{ConcentrationState}, where we see the initial state in red, and the final state in blue.  When $\beta$ is small, the distributions are roughly the same; when $\beta$ is large, the final state has a few peaks that are substantially larger than the rest of the population.  This is the phenomenon of selective amplification.

For both simulations, the initial distribution of concentrations is randomly distributed around $1/N$, and the final distribution is obtained after $4000$ iterations of the wet-dry cycle

\begin{figure}
\includegraphics[width=\columnwidth]{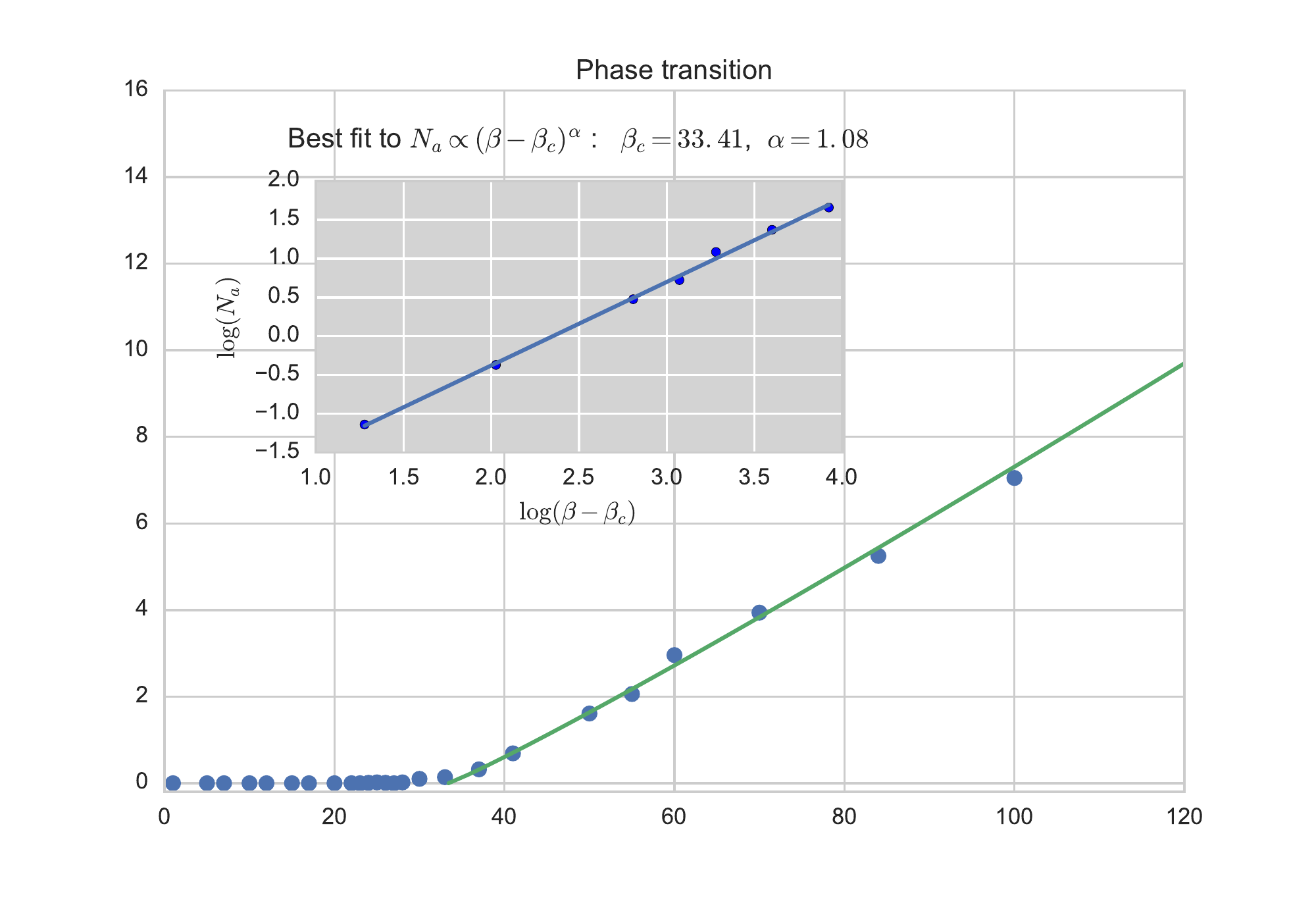}
\caption{The phase transition as $\beta$ is varied, with $N=200$ and $f=0.01$.  The inset graph shows the best fit to a power law for data near the transition, where we see that $\beta_c \approx 33.4$ and the critical exponent $\alpha \approx 1.08$ ($N_a$ commences a linear increase at $\beta_c$).
}\label{PhaseTransition}
\end{figure}

The first metric we consider to capture the essence of selective amplification is to count the number of compounds $N_a$ whose concentration is greater than twice their initial concentration, which we will call the number amplified.  Figure \ref{PhaseTransition} shows how $N_a$ varies with $\beta$.  The simulation to find $\beta_c$ was done with the number of chemical species $N = 1600$, and at each value of $\beta$ 4000 wet-dry cycles were implemented as described above, to arrive at a final concentration profile, from which $N_a$ was computed as the number of species that had at least doubled in concentration.  We checked for a power law at the transition by fitting for $\beta_c$ and $\alpha$, by scanning candidate $\beta_c$ values, performing a fit, and choosing the $\beta_c$ that produced the fit with the lowest standard error, along with the corresponding value of $\alpha$.  This produced $\beta_c \approx 33.4$ and $\alpha = 1.08$, indicating that at $\beta_c$, $N_a$ starts to increase linearly.  Note that the evidence for a power law is not extremely strong, given the limited dynamic range explored in these simulations.

When the system is too hot, the interactions between precipitating pairs become very non-specific, and no transition to sparseness occurs. At lower temperatures (large $\beta$), the differences in binding energies between pairs of compounds become exaggerated, and there begins to be a selection effect. Very small supernatant fractions ($f$) enhance this effect, and at very small $f$ values there appears to be a specific temperature which maximizes the diversity of the amplified subset of compounds. As the temperature is lowered past that point, the system becomes increasingly specific and sparse, resulting in a smaller and smaller set of amplified compounds, along with a lower informational entropy of the chemical distribution.

$N_a$ is somewhat insensitive to small variations, and so we also compute a more continuous metric using the Shannon entropy. If we treat the distribution of concentrations as a probability distribution, we can compute the entropy of that distribution:

\begin{equation}
 S = \sum_i c_i \log( c_i )
\end{equation}

\begin{figure}
\includegraphics[width=\columnwidth]{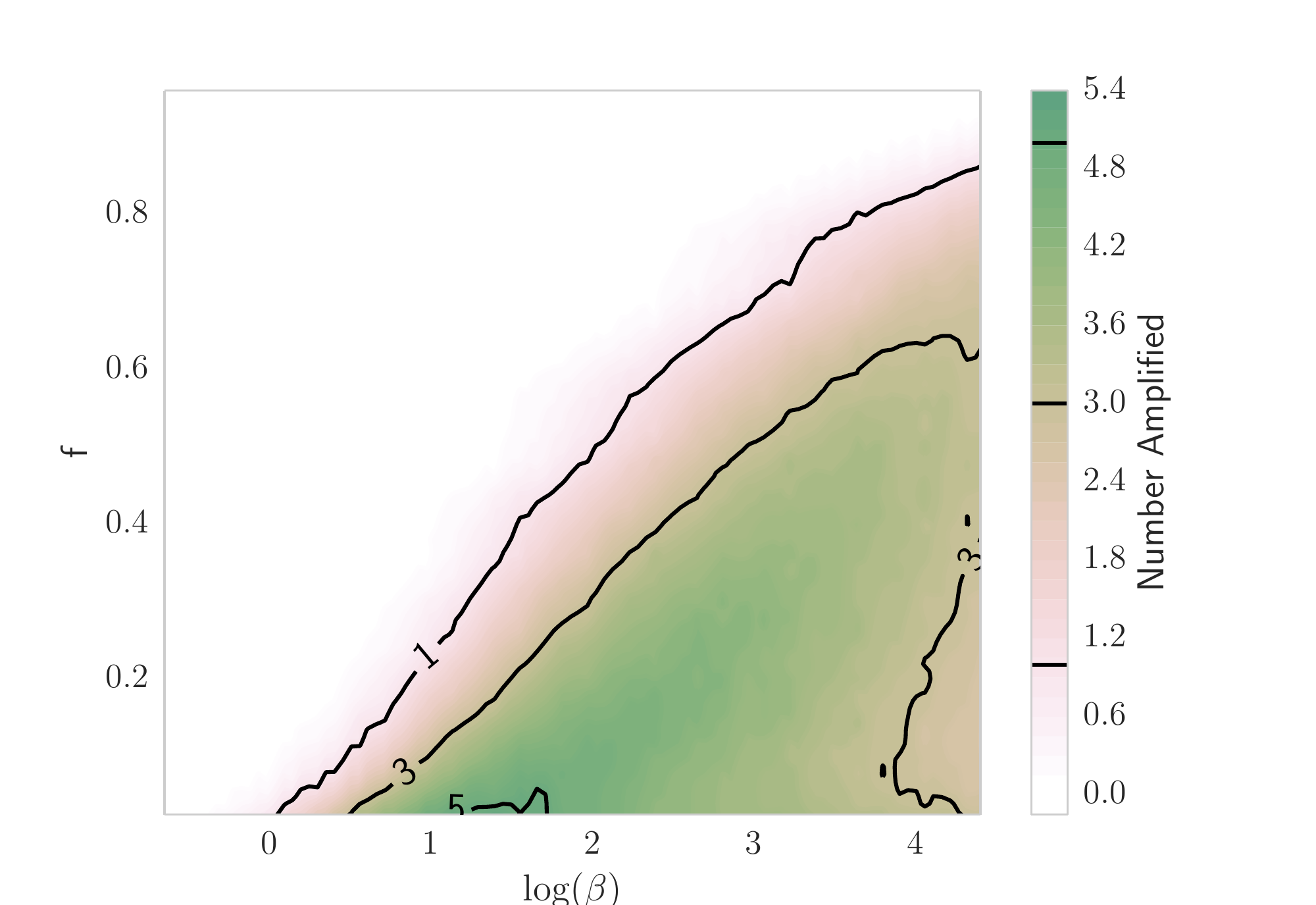}
\includegraphics[width=\columnwidth]{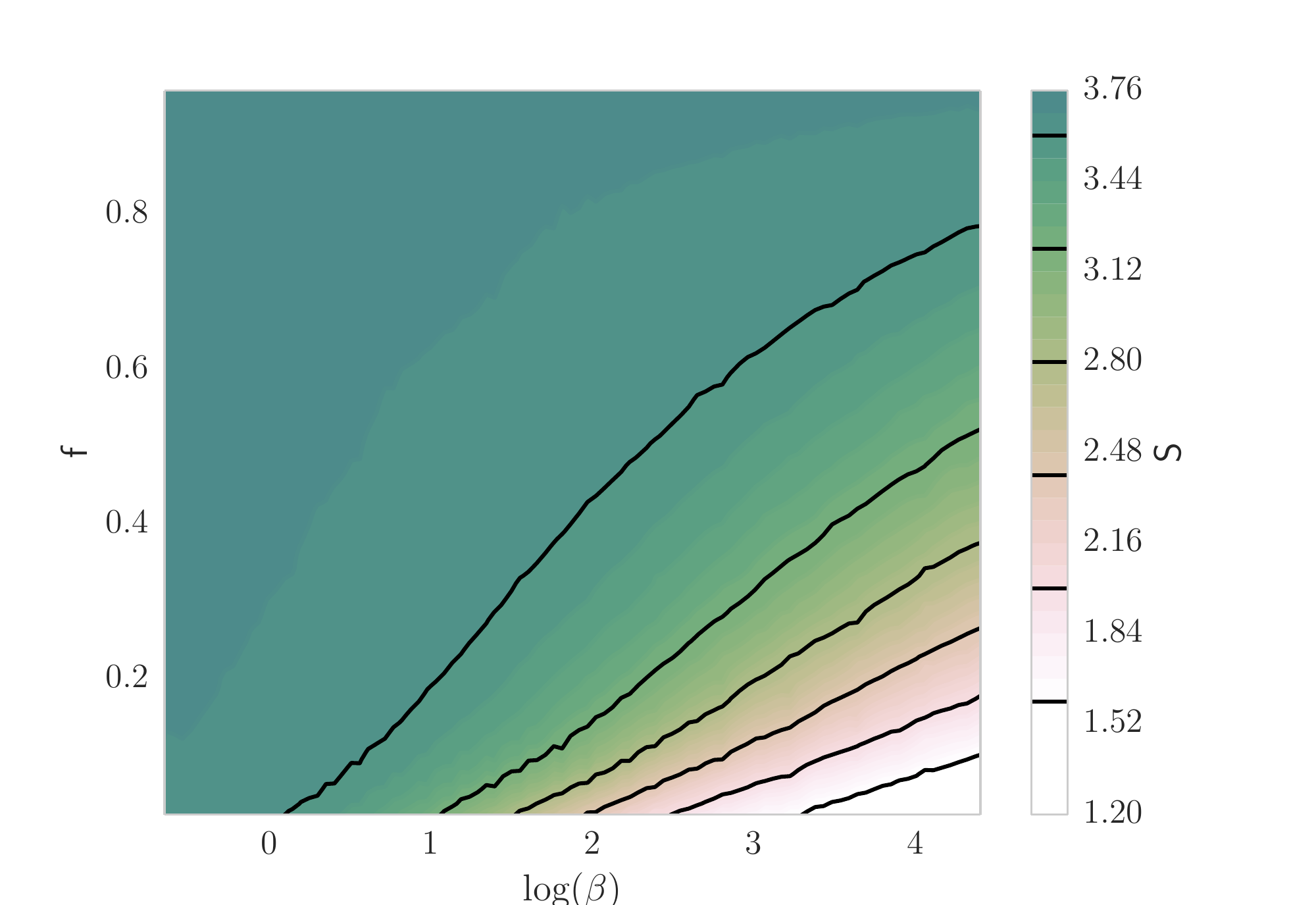}
\caption{Top: Number of compounds amplified to twice their initial concentration as a function of $\beta$ and $f$, averaged over $100$ random $J$ matrices for a system with $N=40$ compounds. The white region in the upper-left corresponds to cases in which there is no amplification taking place. For a cold enough system (larger $\beta$), amplification of individual compounds begins to occur even for very large supernatant fractions, but the set of amplified compounds only becomes diverse when $f$ is also small. Bottom: Information entropy of the steady-state chemical distribution for the same system. Note that even in cases with a single amplified compound, the entropy does not noticeably change compared to the zero amplification case. }\label{PhaseDiagram}
\end{figure}

We calculate a phase diagram of the system given a fixed number of compounds $N=40$. At each value of $\beta$ and $f$, we perform $100$ runs with different random $J$ matrices and compute the average $N_a$ and $S$ (Fig.~\ref{PhaseDiagram}). For these simulations, we use a timestep equal to $0.3\Delta_0$. We performed similar computations for $N=10,20,40,70,100$, and the structure of the phase diagram does not appear to depend strongly on the number of compounds $N$.

\begin{figure}
\includegraphics[width=\columnwidth]{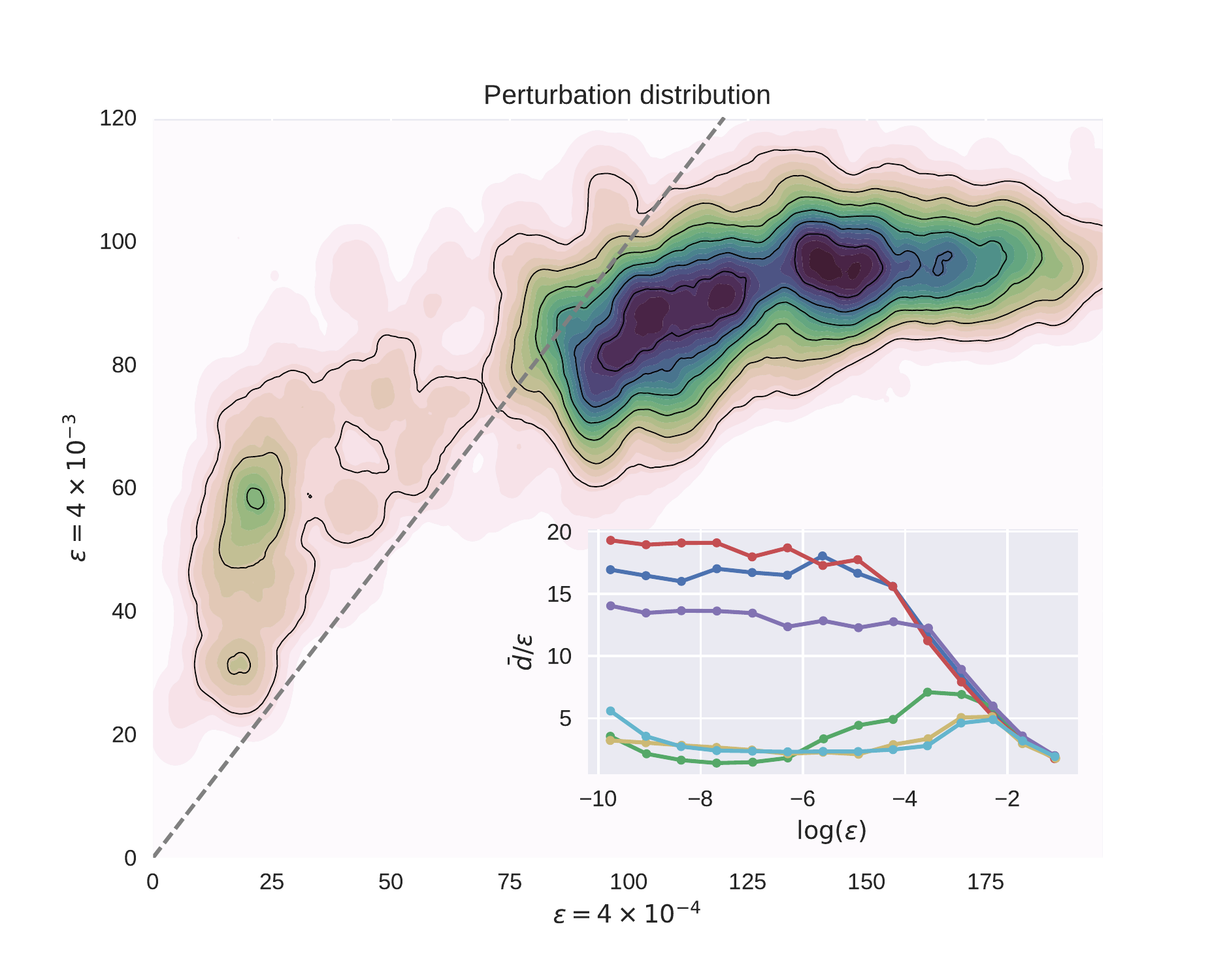}
\caption{Distribution of final state distances as the initial state is perturbed. A few individual perturbation curves are shown in the inset.  ~10,000 perturbation curves were generated and plotted with respect to two coordinates of the function space: the value of $\bar{d}/\epsilon$ at $\epsilon = 4\times10^{-4}$, and the value of $\bar{d}/\epsilon$ at $\epsilon = 4\times10^{-3}$.  The resulting distribution over the 2-dimensional projection of the perturbation curves is illustrated with the contour plot.  The diagonal line is the non-interacting passive chromatography limit; note two separate peaks, one above the diagonal, and a more heavily populated peak below the diagonal.
}\label{responseHistogram}
\end{figure}

\begin{figure}
\includegraphics[width=\columnwidth]{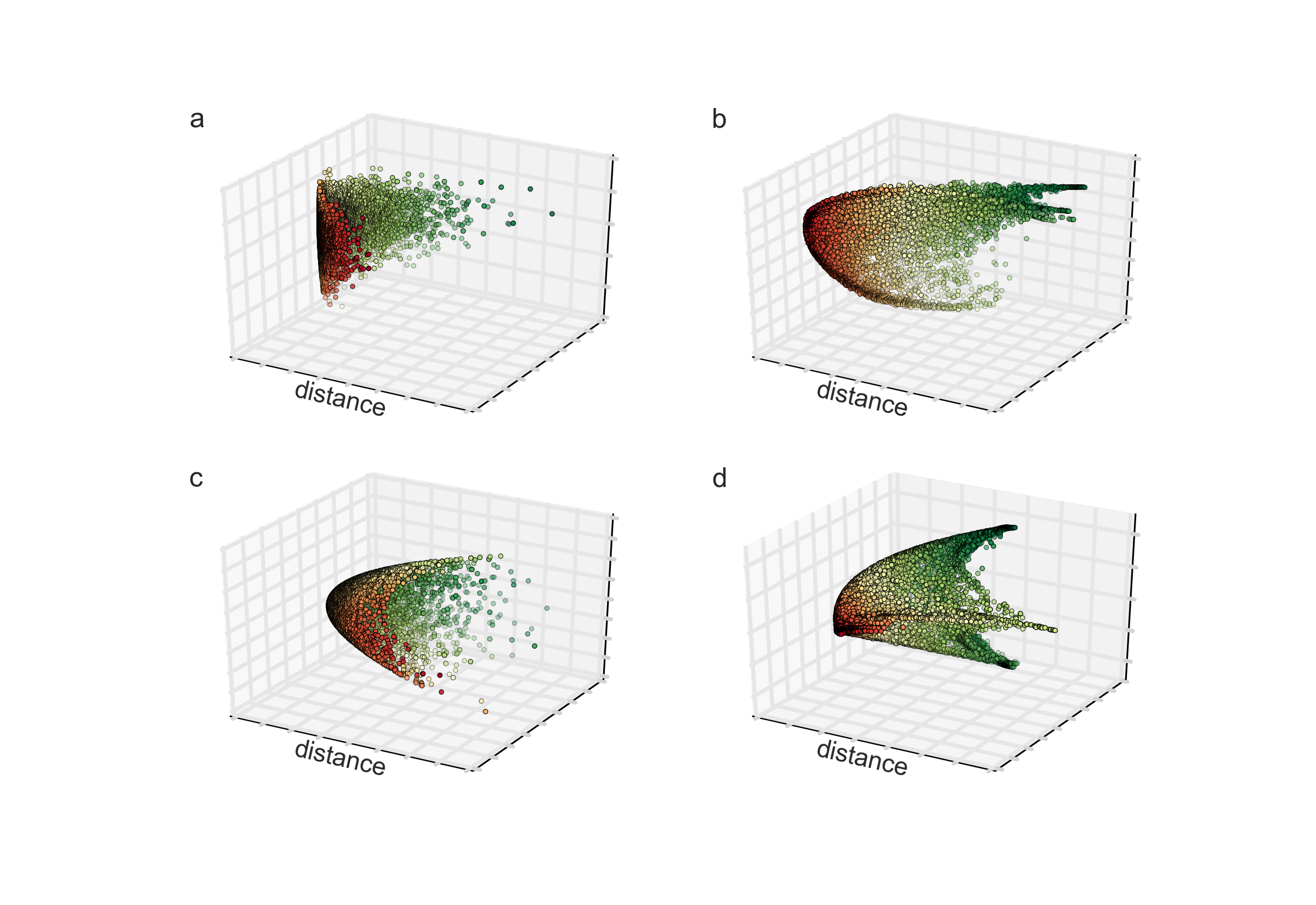}
\caption{Three dimensional scatterplots of perturbed concentration profiles: the first coordinate is distance from the initial environment, and the other two coordinates are the first two principal components of the 200-dimensional concentration, for ~10K samples drawn two different sized perturbations: $\epsilon = 4\times10^{-4}$ and $\epsilon = 4\times10^{-3}$.  The perturbations are made about two different fiducial environments:  (a) and (b) are ensembles of samples around an environment whose distance profile was in the left peak of the previous figure, for small and large perturbations respectively; (c) and (d) are ensembles of samples around an environment whose distance profile was in the right peak of the previous figure, for small and large perturbations respectively.
}\label{responsePCA}
\end{figure}

In general, for a particular environment vector $\vec{e}$ the same specific compounds will be amplified at steady-state. However, what happens when the environment fluctuates or varies?  Response to perturbations should give a sense of the roughness of the selective amplification landscape.
passes through different sets of compounds on the way to reaching its steady-state. This behavior is similar to what one would expect in an evolutionary system with competitive exclusion\cite{hardin1960competitive} --- that is to say, in such cases (e.g. in the absence of pairwise interactions) the only stable attractor at long times is the global fitness maximum. In order to compare this system with the results from GARD\cite{Segre:1998dj} and catalytic networks, we wish to measure the size of the potential evolutionary space of the intrinsic dynamics (that is, the chemical properties of the compounds in the system), not just the system's response to a single particular environment. That is to say, we want to measure the total number of distinct ways that the system can adapt.

If the wet-dry cycling process were just passive chromatography with no pairwise interactions between components, we would expect that a perturbation of the environment $\vec{e} \rightarrow \vec{e} + \vec{\epsilon}$ would create a proportional change in the concentration profile in steady state $\vec{c} \rightarrow \vec{c} + \lambda \vec{\epsilon}$. Depending on the strength of the affinity of the column, $\lambda$ could be large or small (with large affinities corresponding to small responses), but the response should be linear.

For our wet-dry cycle with the nonlinear pairwise interactions given by Eq. \ref{Model}, the effect of perturbations will not necessarily be linear.   To observe the effect of perturbations, for a fixed $\textbf{J}$ ($N=200, \beta=100$), we vary the size of the perturbation $\vec{\epsilon}$ and measure the difference between the original state $c$ and the final state $c + d(\vec{\epsilon})$, averaging over 16 random perturbations of the environment vector to get $\bar{d}(\epsilon)$. Here, the components of the perturbations were each chosen from a uniform distribution of $[-\epsilon,\epsilon]$, and if a perturbation caused one of the concentrations to go to negative, the offending concentration was clamped to zero, and all concentrations renormalized.  The inset to Fig.~\ref{responseHistogram} displays the perturbation response curves for six different $\textbf{J}$'s. While many of the response curves from the selective amplification model have constant $d(\epsilon)/\epsilon$ (at least up to a point of saturation), as one might expect for passive chromatography, there appears to be a family of response curves where initially the response becomes proportionately weaker as $|\vec{\epsilon}|$ increases, and then becomes significantly stronger, examples seen in the bottom perturbation response curves of the inset graph of Fig.~\ref{responseHistogram}.  Also of some surprise is that there is quite a large variation in the response curve observed, depending on the particulars of $\textbf{J}$ and of the initial environment vector $\vec{e}$.

To explore the full range of perturbation responses, we repeated the measurement of perturbation of responses (as shown in the inset graph of Fig.~\ref{responseHistogram}) 900 times, for 900 different randomly chosen $\textbf{J}$.  We visualize the distribution of perturbation response curves by projecting this space of functions onto two dimensions, by looking at two values, the value of $\bar{d}/\epsilon$ at $\epsilon = 4\times10^{-4}$, and the value of $\bar{d}/\epsilon$ at $\epsilon = 4\times10^{-3}$.  The contour plot of Fig.~\ref{responseHistogram} shows the distribution of perturbation responses projected onto these two coordinates.  We see that there seem to be two major clusters of results --- one set in which the sensitivity to perturbation $\bar{d}(\epsilon)/\epsilon$ is relatively low but highly nonlinear with the strength of the perturbation (the peak over the $y=x$ diagonal), and another set in which the sensitivity to perturbation $\bar{d}(\epsilon)/\epsilon$ is quite high but relatively constant  (the large, broad peak to the right of the $y=x$ line).

Finally, we look at the effect of perturbations on the concentration profiles themselves, for two representative cases in each of the two peaks just described, {\em i.e.}, for two different chemistries with two different $\textbf{J}$'s.  We sample 10,000 perturbations of size $\epsilon$ (the sampling procedure as before), compute the principal components of the resulting final concentration profiles, and look at the projection of the profiles into the three dimensional space consisting of the distance of the final profile from the original profile along with the first two principal components.  This was performed for two different representative chemistries in each of the two peaks of Fig.~\ref{responseHistogram}, for two different perturbations sizes, $\epsilon = 4 \times 10^{-4}$ and $\epsilon = 4 \times 10^{-3}$.  The results are shown in Fig.~\ref{responsePCA}.  We see that for small perturbations, Fig.~\ref{responsePCA} (a) and (c), the response takes a simple form parabolic in the first principal component, and parabolic in the first two principal components.  Larger perturbations, represented in Fig.~\ref{responsePCA} (b) and (d), create more complex responses.  The small perturbation alters the distribution by changing the size of the concentration peaks; the large perturbation alters the identity of the peaks.

\section{Implications for the origin of life}

The onset of evolution is taken to be a crossover point from a domain in which every stage of synthesis must be established as environmentally present, into a domain in which the system's own exploration and population dynamics takes over and can discover and optimize novel chemistries and synthesis pathways. This creates an apparent difficulty, in that systems that look like modern sequence-based evolution require quite a lot of scaffolding in order to support themselves against entropic effects. This difficulty is seen as one of the major stumbling blocks of the RNA world narrative for the origin of life.\cite{LeslieE:2010jz}

However, it may be that the onset of evolution is not so binary, but that as a precursor to contemporary evolutionary processes based on variations in informational polymers, we would see a sequence of different types of increasingly self-referential selection --- starting with passive chromatography and fractionation that arises just from energy minimization, with a gradual transition to selection which depends on the pairwise interactions and functional character of the components of the system.  In that case, rather than evolution performing a directed search through a space of polymeric possibilities, we would have a process which simply amplifies the latent properties contained within the chemical library given to it by environmental processes and chemical cycles, but one which does so in a way that focuses on things that are in turn chemically relevant to the compounds that end up being amplified (in a self-consistent way).

In this scenario, information is contained not in polymeric identity, but rather in the identity of self-amplified populations.  We have shown how there can be substantial variations in these populations with small perturbations,  and it is easy to envision  perturbations caused, e.g. by spatial separations of sequestration processes.

The wet-dry mechanism we discuss here is one possibility by which this form of weak adaptation can take place, by way of a positive feedback mechanism which refines a diverse low-concentration chemical library into a sparse, high concentration set of particular compounds selected on based on their functional properties (in this case, their ability to drive co-precipitation). This sort of mechanism is potentially quite generic, requiring only some form of cyclic fractionation process combined with inflow and outflow.  Besides occurring in wet-dry cycles, this type of process could occur in repeated , cyclic transport through porous media combined with deposition and dissolution, freeze-thaw cycles, or in boundary structures such as surface films or viscosity gradients within tars. Furthermore, because this mechanism operates via pairwise interactions and selectivity, it may provide a natural bridge towards the evolution of sequence-based replication. If the compounds which pair during repeated sequestration can themselves begin to have structural complexity, e.g. by taking on covalently linked backbones, then base-pairing conjugate sequences would tend to be strongly selected for due to having very strong mutual affinity along with being able to occasionally catalyze their own replication from monomers. 

Selective amplification through repeated sequestration is not necessarily the only way primordial evolution without polymeric information could take place --- we have presented a demonstration that the possibility exists, not a detailed argument that this particular process was how it happened on the Hadean Earth.  At the same time however, it may be interesting and worthwhile to investigate more deeply the effect of these sorts of fractionation processes on complex chemical systems and self-assembly processes associated with the prebiotic Earth. Miller-Urey tars and formose tars in particular may be good vehicles for further exploration along these lines, as they represent sources of diverse chemical libraries whose extreme diversity has in some sense posed a limiting factor to their ability to create the more rarefied and specific chemical networks associated with biological chemistry.

\section{Acknowledgements}

We would like to acknowledge Bruce Damer and David Deamer for discussions about experimental realizations of this mechanism. This project was partially supported by the ELSI Origins Network (EON), which is supported by a grant from the John Templeton Foundation. The opinions expressed in this publication are those of the author(s) and do not necessarily reflect the views of the John Templeton Foundation.

\bibliography{references}

\begin{thebibliography}{8}%
\makeatletter
\providecommand \@ifxundefined [1]{%
 \@ifx{#1\undefined}
}%
\providecommand \@ifnum [1]{%
 \ifnum #1\expandafter \@firstoftwo
 \else \expandafter \@secondoftwo
 \fi
}%
\providecommand \@ifx [1]{%
 \ifx #1\expandafter \@firstoftwo
 \else \expandafter \@secondoftwo
 \fi
}%
\providecommand \natexlab [1]{#1}%
\providecommand \enquote  [1]{``#1''}%
\providecommand \bibnamefont  [1]{#1}%
\providecommand \bibfnamefont [1]{#1}%
\providecommand \citenamefont [1]{#1}%
\providecommand \href@noop [0]{\@secondoftwo}%
\providecommand \href [0]{\begingroup \@sanitize@url \@href}%
\providecommand \@href[1]{\@@startlink{#1}\@@href}%
\providecommand \@@href[1]{\endgroup#1\@@endlink}%
\providecommand \@sanitize@url [0]{\catcode `\\12\catcode `\$12\catcode
  `\&12\catcode `\#12\catcode `\^12\catcode `\_12\catcode `\%12\relax}%
\providecommand \@@startlink[1]{}%
\providecommand \@@endlink[0]{}%
\providecommand \url  [0]{\begingroup\@sanitize@url \@url }%
\providecommand \@url [1]{\endgroup\@href {#1}{\urlprefix }}%
\providecommand \urlprefix  [0]{URL }%
\providecommand \Eprint [0]{\href }%
\providecommand \doibase [0]{http://dx.doi.org/}%
\providecommand \selectlanguage [0]{\@gobble}%
\providecommand \bibinfo  [0]{\@secondoftwo}%
\providecommand \bibfield  [0]{\@secondoftwo}%
\providecommand \translation [1]{[#1]}%
\providecommand \BibitemOpen [0]{}%
\providecommand \bibitemStop [0]{}%
\providecommand \bibitemNoStop [0]{.\EOS\space}%
\providecommand \EOS [0]{\spacefactor3000\relax}%
\providecommand \BibitemShut  [1]{\csname bibitem#1\endcsname}%
\let\auto@bib@innerbib\@empty
\bibitem [{\citenamefont {Benner}\ \emph {et~al.}(2010)\citenamefont {Benner},
  \citenamefont {Kim}, \citenamefont {Kim},\ and\ \citenamefont
  {Ricardo}}]{benner2010planetary}%
  \BibitemOpen
  \bibfield  {author} {\bibinfo {author} {\bibfnamefont {S.~A.}\ \bibnamefont
  {Benner}}, \bibinfo {author} {\bibfnamefont {H.-J.}\ \bibnamefont {Kim}},
  \bibinfo {author} {\bibfnamefont {M.-J.}\ \bibnamefont {Kim}}, \ and\
  \bibinfo {author} {\bibfnamefont {A.}~\bibnamefont {Ricardo}},\ }\href@noop
  {} {\bibfield  {journal} {\bibinfo  {journal} {Cold Spring Harbor
  perspectives in biology}\ }\textbf {\bibinfo {volume} {2}},\ \bibinfo {pages}
  {a003467} (\bibinfo {year} {2010})}\BibitemShut {NoStop}%
\bibitem [{\citenamefont {Mamajanov}\ \emph {et~al.}(2014)\citenamefont
  {Mamajanov}, \citenamefont {MacDonald}, \citenamefont {Ying}, \citenamefont
  {Duncanson}, \citenamefont {Dowdy}, \citenamefont {Walker}, \citenamefont
  {Engelhart}, \citenamefont {Fernández}, \citenamefont {Grover},
  \citenamefont {Hud} \emph {et~al.}}]{mamajanov2014ester}%
  \BibitemOpen
  \bibfield  {author} {\bibinfo {author} {\bibfnamefont {I.}~\bibnamefont
  {Mamajanov}}, \bibinfo {author} {\bibfnamefont {P.~J.}\ \bibnamefont
  {MacDonald}}, \bibinfo {author} {\bibfnamefont {J.}~\bibnamefont {Ying}},
  \bibinfo {author} {\bibfnamefont {D.~M.}\ \bibnamefont {Duncanson}}, \bibinfo
  {author} {\bibfnamefont {G.~R.}\ \bibnamefont {Dowdy}}, \bibinfo {author}
  {\bibfnamefont {C.~A.}\ \bibnamefont {Walker}}, \bibinfo {author}
  {\bibfnamefont {A.~E.}\ \bibnamefont {Engelhart}}, \bibinfo {author}
  {\bibfnamefont {F.~M.}\ \bibnamefont {Fernández}}, \bibinfo {author}
  {\bibfnamefont {M.~A.}\ \bibnamefont {Grover}}, \bibinfo {author}
  {\bibfnamefont {N.~V.}\ \bibnamefont {Hud}},  \emph {et~al.},\ }\href@noop {}
  {\bibfield  {journal} {\bibinfo  {journal} {Macromolecules}\ }\textbf
  {\bibinfo {volume} {47}},\ \bibinfo {pages} {1334} (\bibinfo {year}
  {2014})}\BibitemShut {NoStop}%
\bibitem [{\citenamefont {Rodriguez-Garcia}\ \emph {et~al.}(2015)\citenamefont
  {Rodriguez-Garcia}, \citenamefont {Surman}, \citenamefont {Cooper},
  \citenamefont {Su{\'a}rez-Marina}, \citenamefont {Hosni}, \citenamefont
  {Lee},\ and\ \citenamefont {Cronin}}]{rodriguez2015formation}%
  \BibitemOpen
  \bibfield  {author} {\bibinfo {author} {\bibfnamefont {M.}~\bibnamefont
  {Rodriguez-Garcia}}, \bibinfo {author} {\bibfnamefont {A.~J.}\ \bibnamefont
  {Surman}}, \bibinfo {author} {\bibfnamefont {G.~J.}\ \bibnamefont {Cooper}},
  \bibinfo {author} {\bibfnamefont {I.}~\bibnamefont {Su{\'a}rez-Marina}},
  \bibinfo {author} {\bibfnamefont {Z.}~\bibnamefont {Hosni}}, \bibinfo
  {author} {\bibfnamefont {M.~P.}\ \bibnamefont {Lee}}, \ and\ \bibinfo
  {author} {\bibfnamefont {L.}~\bibnamefont {Cronin}},\ }\href@noop {}
  {\bibfield  {journal} {\bibinfo  {journal} {Nature communications}\ }\textbf
  {\bibinfo {volume} {6}} (\bibinfo {year} {2015})}\BibitemShut {NoStop}%
\bibitem [{\citenamefont {Seto}\ and\ \citenamefont
  {Whitesides}(1993)}]{seto1993molecular}%
  \BibitemOpen
  \bibfield  {author} {\bibinfo {author} {\bibfnamefont {C.~T.}\ \bibnamefont
  {Seto}}\ and\ \bibinfo {author} {\bibfnamefont {G.~M.}\ \bibnamefont
  {Whitesides}},\ }\href@noop {} {\bibfield  {journal} {\bibinfo  {journal}
  {Journal of the American Chemical Society}\ }\textbf {\bibinfo {volume}
  {115}},\ \bibinfo {pages} {905} (\bibinfo {year} {1993})}\BibitemShut
  {NoStop}%
\bibitem [{\citenamefont {Cafferty}\ \emph {et~al.}(2013)\citenamefont
  {Cafferty}, \citenamefont {Gállego}, \citenamefont {Chen}, \citenamefont
  {Farley}, \citenamefont {Eritja},\ and\ \citenamefont
  {Hud}}]{cafferty2013efficient}%
  \BibitemOpen
  \bibfield  {author} {\bibinfo {author} {\bibfnamefont {B.~J.}\ \bibnamefont
  {Cafferty}}, \bibinfo {author} {\bibfnamefont {I.}~\bibnamefont {Gállego}},
  \bibinfo {author} {\bibfnamefont {M.~C.}\ \bibnamefont {Chen}}, \bibinfo
  {author} {\bibfnamefont {K.~I.}\ \bibnamefont {Farley}}, \bibinfo {author}
  {\bibfnamefont {R.}~\bibnamefont {Eritja}}, \ and\ \bibinfo {author}
  {\bibfnamefont {N.~V.}\ \bibnamefont {Hud}},\ }\href@noop {} {\bibfield
  {journal} {\bibinfo  {journal} {Journal of the American Chemical Society}\
  }\textbf {\bibinfo {volume} {135}},\ \bibinfo {pages} {2447} (\bibinfo {year}
  {2013})}\BibitemShut {NoStop}%
\bibitem [{\citenamefont {Hardin}\ \emph {et~al.}(1960)\citenamefont {Hardin}
  \emph {et~al.}}]{hardin1960competitive}%
  \BibitemOpen
  \bibfield  {author} {\bibinfo {author} {\bibfnamefont {G.}~\bibnamefont
  {Hardin}} \emph {et~al.},\ }\href@noop {} {\bibfield  {journal} {\bibinfo
  {journal} {Science}\ }\textbf {\bibinfo {volume} {131}},\ \bibinfo {pages}
  {1292} (\bibinfo {year} {1960})}\BibitemShut {NoStop}%
\bibitem [{\citenamefont {Segr{\'e}}\ \emph {et~al.}(1998)\citenamefont
  {Segr{\'e}}, \citenamefont {Lancet}, \citenamefont {Kedem},\ and\
  \citenamefont {Pilpel}}]{Segre:1998dj}%
  \BibitemOpen
  \bibfield  {author} {\bibinfo {author} {\bibfnamefont {D.}~\bibnamefont
  {Segr{\'e}}}, \bibinfo {author} {\bibfnamefont {D.}~\bibnamefont {Lancet}},
  \bibinfo {author} {\bibfnamefont {O.}~\bibnamefont {Kedem}}, \ and\ \bibinfo
  {author} {\bibfnamefont {Y.}~\bibnamefont {Pilpel}},\ }\href {\doibase
  10.1023/A:1006583712886} {\bibfield  {journal} {\bibinfo  {journal} {Origins
  of Life and Evolution of the Biosphere}\ }\textbf {\bibinfo {volume} {28}},\
  \bibinfo {pages} {501} (\bibinfo {year} {1998})}\BibitemShut {NoStop}%
\bibitem [{\citenamefont {Leslie~E}(2010)}]{LeslieE:2010jz}%
  \BibitemOpen
  \bibfield  {author} {\bibinfo {author} {\bibfnamefont {O.}~\bibnamefont
  {Leslie~E}},\ }\href {\doibase 10.1080/10409230490460765} {\bibfield
  {journal} {\bibinfo  {journal} {Critical Reviews in Biochemistry and
  Molecular Biology}\ }\textbf {\bibinfo {volume} {39}},\ \bibinfo {pages} {99}
  (\bibinfo {year} {2010})}\BibitemShut {NoStop}%
\end{thebibliography}%

\end{document}